 \documentclass[11pt,floatfix,notitlepage]{article}
 \usepackage[english]{babel}
 \usepackage{amsmath,mathrsfs,amssymb,esint,cancel}
 \usepackage{pst-eps,auto-pst-pdf}
 \usepackage{caption2,graphics,placeins,verbatim}
 \usepackage[totalwidth=500pt, totalheight=650pt]{geometry}
 \usepackage{etex}
 \usepackage{CJKutf8}
 \usepackage[unicode]{hyperref}
 \usepackage{indentfirst}
 \usepackage{setspace}
 \newcommand{\beq}[1]{\begin{equation}\label{#1}}
 \newcommand{\eeq}{\end{equation}}
 \newcommand{\bea}[1]{\begin{eqnarray}\label{#1}}
 \newcommand{\eea}{\end{eqnarray}}

\begin{document}
\begin{CJK}{UTF8}{gbsn}
\CJKtilde

\title{Scale-free power spectrums in the delayed cosmology}
\author{Shi-hui Yang and Ding-fang Zeng\\
Theoretical Physics Division, College of Applied Sciences\\
Beijing University of Technology, Beijing, China
\\
shihuiyang@icloud.com and dfzeng@bjut.edu.cn}

\date{\today}
\maketitle

\abstract{The delayed cosmology [JCAP 02(2012)046] assumes that the evolution of geometries is delayed relative to that of matter and/or energies. This idea allows inflation occur without inflaton fields or vacuum energies of any kind as drivings. We considered the production and evolution of primordial perturbations in this model. The result indicate that, with delaying, we could get a nearly scale-free power spectrum consistent with observations starting from a totally radiation dominated early universe.
}

\tableofcontents

\section{Introduction}
The scenario of inflation \cite{Guth, Linde} developed since 1980s may be the most successful part of modern cosmology \cite{LLReview, BaumannReview}. It solves the horizon, flatness and other  unavoidable problems of old cosmology very successfully. Most importantly, it provides an elegant mechanism for the production of primordial seeds of structures of the universe \cite{LLReview, WMAP}. On March 2014, the BICEP2 team announced their detection of primordial gravitational waves through the B-mode polarization \cite{BICEP2} of cosmological microwave background radiation, which is expected to be a strong evidence for the scenario of inflation. However, the joint analysis of BICEP2 and Planck Collaborations in Febrary 2015 indicates that, this phenomenon is mostly due to the effects of galaxy dusts\cite{BICEP2_PLANCK}. 
As a result, the settling down of the occurrence of inflation becomes an open question again. Nevertheless,  the scenario of inflation still remains as the most appealing idea for the early history of the universe.

In most of the existing models, see for examples \cite{LLReview,BaumannReview}, inflations during the early universe is implemented with the aid of some one or more scalar fields named inflaton. As far as we know, all these kinds of inflation models violate the strong energy conditions, some of them require specifically designed mechanism to protect the universe from eternal inflation, and some others of them require artificial choosing of initial values for the inflaton fields. What's more, in the standard model of elementary particles, we have not yet found any scalar fields could properly play the role of inflaton. 

\begin{figure*}
\setcaptionwidth{0.85\textwidth}
\begin{center}
\includegraphics[scale=0.8]{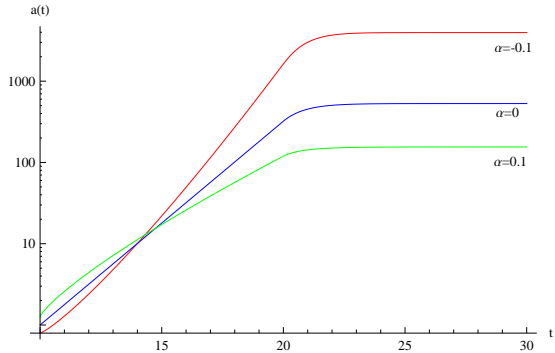}
\caption{By introducing a delay to the source term of Friedmann equation, inflation is obtained in the absence of extra scalar field, and is accompanied with a smooth graceful exit for a wide range of $\alpha$. In this figure the delay parameter is chosen as $\tau=10t_{pl}$. The inflation lasts in the period $\tau<t<2\tau$.}
\label{inflationDelay}
\end{center}
\end{figure*}

Noticing that in many natural phenomenas, the responses of systems are usually delayed relative to their driving forces, Choudhury et al. proposed \cite{delayedCosmology} that, in the Friedmann equation, similar delaying effects may also take place,
\beq{}
\big[\frac{\dot a(t)}{a(t)}\big]^2=\frac{1}{3}\rho(t-\tau).
\label{friedmannDelayed}
\eeq
Here $a$, $\rho$, $t$ and overdot denote the scale factor, energy density, cosmic time and derivatives respect to $t$ respectively, $\tau$ is the delaying. With this assumption and some rather general initial conditions, Choudhury et al found that an early inflation and elegant exit from it would take place very naturally. Of course, to prevent possible contradictions between the known observations and predictions following from Eq. \eqref{friedmannDelayed}, the parameter $\tau$ should not be too large. Choudhury et al. find that under the pre-inflation assumption $a_{0<t<\tau}=t^\alpha$, the resulting evolution of scale factors has the form
\beq{}
a(t) =\Bigg\{\begin{array}{l}
{\tau^\alpha}\exp\Big(H_i \frac{(t-\tau)^{1-\frac{3}{2}(1+w)\alpha}
}{1-\frac{3}{2}(1 + w)\alpha}\Big),~\tau<t<2\tau
\\
\mathrm{graceful~exit},~t<\infty
\end{array}
\label{aoftDelayed}
\eeq
Obviously, as long as $(1+w)\alpha<\frac{2}{3}$, the universe will experience accelerations in appropriate later periods, see Figure \ref{inflationDelay} for illustrations. Following Ref. \cite{delayedCosmology}, we will call this idea as delayed cosmology, or sometimes, delayed inflation. We comment here that, more general form of pre-inflation scale factors are also allowed and it is in fact counter-part of potentials in the scalar field driving inflation models.

This is an ingenious idea for the mechanism of inflation. It avoids the introduction of "professional'' fields which is not in the known lists of particle physics as driving forces, and
solves the exit problem gracefully. Certainly, as a new kind of inflation mechanism, enabling the exponential growth of scale factors is not adequate. The more important question is, could this mechanism provide the seeds of structures for the later universe. As it is well known that, in the conventional models, the quantum fluctuations of inflaton fields after being pulled out of horizon by the accelerating expansion, it became the seeds of structures for late time evolutions. For a mechanism without inflaton fields as the delayed cosmology, we could only depend on the quantum fluctuations of pre-inflation cosmic-contents (radiation/matter) to provide such seeds of structures. The purpose of this paper is to consider the evolution of perturbations in the delayed inflation and calculate the power spectrum of primordial fluctuations.

The organization of this paper is as follow, this section is a brief introduction to the basic ideas of delayed cosmology. The next section calculates the power spectrums of primordial perturbations produced in the delayed inflation, includes both scalar and tensor modes. The next next section numerically evolves the primordial fluctuations in the late time universe with delays and obtain the power spectrum of matter distributions today. The last section is the conclusion.

\section{The power spectrum of perturbations in the delayed inflation}
\label{sectionPowerSpectrum}
\subsection{The equation of motion for perturbations}
The basic idea of delayed cosmology/inflation is revising the Friedman equation so that the growth of the cosmic scale factor is delayed relative to the evolution of energy-densities. When we consider the questions of perturbation and structure formation, it is very natural to generalize this idea to the full Einstein equation, so that
\beq{}
{G_{\mu \nu }}\left( {\vec x,t} \right) = {T_{\mu \nu }}\left( {\vec x,t - \tau } \right).
\label{delayedEinsteinEq}
\eeq
It is worthwile pointing out that, this delaying generalisation of Einstein equation breaks general covariance. The delayed Friedmann equation \eqref{friedmannDelayed}, if following from some quantum theory of gravity, requires this feature unavoidably. Our generalisation may be the simplest one. It allows us to borrow perturbation techniques from Einstein theory to maximum degree.

According to the theory of cosmological perturbations \cite{CPT, CIP, ModernCosmology}, in the Newton + gravitational wave gauge, we could write the perturbed cosmic metric as
\beq{}
ds^2=-(1-2\Phi)dt^2+a^2[\delta_{ij}(1+2\Phi)+h_{ij}]dx^idx^j
\eeq
where $h_{ij}$ takes the form of
\beq{}
h_{ij}=\left(\begin{array}{ccc}
h_+&h_\times&0\\
h_\times&-h_+&0\\
0&0&0
\end{array}\right)
\eeq
Substituting this perturbed metric into the delayed Einstein equation \eqref{delayedEinsteinEq}, and taking first order approximations, we will get
\beq{}
\nabla^2 \Phi  - 3\mathcal{H}\Phi '  - 3{\mathcal{H}^2}\Phi  = \frac{1}{2}{a^2}\delta {\rho _\tau }
\label{eqAScalarPerturb}
\eeq
\beq{}
{D_i}\left( {\Phi  '+ \mathcal{H}\Phi } \right) =  - \frac{1}{2}{a^2}\left( {{\rho _\tau } + {p_\tau }} \right){\nabla _i}V
\label{eqBScalarPerturb}
\eeq
\beq{}
\Phi'' + 3\mathcal{H} \Phi' + \left( {2{\mathcal{H}'} + {\mathcal{H}^2}} \right)\Phi  = \frac{1}{2}{a^2}\delta {p _\tau }
\label{eqCScalarPerturb}
\eeq
for scalar perturbation and
\beq{}
h_\lambda'' + 2\mathcal{H}h_\lambda' + {k^2}{h_\lambda } = 0
\label{eqTensorPerturb}
\eeq
for tensor perturbations. The subscripts $\tau$ on the right hand side of these equation means delaying, i.e. $\{\rho_\tau,p_\tau,\cdots\}\equiv\{\rho(t-\tau), p(t-\tau),\cdots\}$. While the prime $'$ symbols appearing on the left hand side represents derivatives respect to the conformal time $\frac{d}{d\eta}=a\frac{d}{dt}$, so $\mathcal{H}\equiv\frac{1}{a}\frac{da}{d\tau}=\frac{da}{dt}$.

For general non adiabatic perturbations, $\delta p=c_s^2\delta\rho+\delta p_{nad}$. The resulting equation array \eqref{eqAScalarPerturb}-\eqref{eqCScalarPerturb}  will be very difficult to process. However, for the physically more relevant adiabatic perturbations, the source term in these equations can be eliminated immediately through simple combinations. The resulting equation reads
\beq{}
\Phi''+ 3\mathcal{H}\left( {1 + c_s^2} \right)\Phi'  + c_s^2{k^2}\Phi  + [2\mathcal{H}' +(1+3c_s^2)\mathcal{H}^2]\Phi  = 0,
\label{eqHomoScalarPerturb}
\eeq
where $c_s^2\equiv\frac{\partial p}{\partial\rho}|_\mathrm{adiabatic}=w$. Obviously, delaying effects enter these equations only through the coefficient functions such as $\mathcal{H}$ and $\mathcal{H}'$ et al. Physically, these equations have no difference relative to those in conventional scalar field driven inflation \cite{LLReview}. Techniquely, since the Newton potential $\Phi$ in these equations have no ``entanglement'' with other fields like those in scalar field driven inflations. We have to calculate its power spectrum directly. This is different from the case in scalar field driven inflation theories, where one usually get the spectrum of  $\Phi$ through that of the inflaton field which is more easy to treat. 

\subsection{The power spectrum of perturbations, exponential inflation}
Now, let us follow the standard method of inflationary cosmology, and consider the quantization of perturbations in the delayed inflation. For tensor perturbation, the equation of motion and power spectrum expressions are completely of the same form as they are in scalar field driving models \cite{spectrum}. The only point which should be emphasized here is that, in quantization of $h$, we have to make the transformation
\beq{}
\tilde h\equiv \frac{1}{\sqrt 2}ah=
\hat{a}_{\vec k}v(\vec k,\eta)+\hat{a}_{\vec k}^{\dag}v^*(\vec k,\eta)
\label{hhTilde}
\eeq
so that
\beq{}
v''+(k^2-\frac{a''}{a})v=0.
\label{eqTildeh}
\eeq
where $\prime$ still represents derivatives with respect to conformal time $\eta$.
For exponential inflation (in delayed inflation, this means $\alpha=0$), the Hubble parameter $H$ is a constant. 
\beq{}
\eta  = \int_{a_e}^a\frac{da}{Ha^2} =-\frac{1}{aH}\Big|_{a_e}^a~\Rightarrow~\frac{a''}{a}=\frac{2}{\eta^2}
\label{etaDefinition}
\eeq
where $a_e$ is the scale factor as inflation ends. In this case, equation \eqref{eqTildeh} can be solved analytically
\beq{}
v(\vec{k},\eta)=\frac{1}{\sqrt{2k}}e^{-ik\eta}(1-\frac{i}{k\eta})
\eeq
At the beginning of inflation, $a\rightarrow0$, $\eta\rightarrow-\infty$, $-k\eta\gg1$, the purturbation is inside the horizon $v(\vec{k},\eta)=\frac{1}{\sqrt{2k}}e^{-ik\eta}$. While at the ending of inflation, $a\rightarrow a_e$, $\eta\rightarrow0$, $k\eta\ll1$. In this case, the perturbation is stretched outside the horizon $v(\vec{k},\eta)=-\frac{1}{\sqrt{2k}}\frac{ie^{-ik\eta}}{k\eta}$.
The variance or power spectrum of the quantized $\tilde h$ take its value at this time
\beq{}
\hspace{-5mm}P_{\tilde{h}}(k)\equiv|v|^2_{-k\eta\ll1}=\frac{a^2H^2}{2k^3}|_{a=a_e}
\label{hoSpectra}
\eeq
According to the definition \eqref{hhTilde},
the power spectrum of $h$ could be calculated from that of $\tilde{h}$,
\beq{}
P_h\equiv\frac{2}{a^2_e}P_{\tilde{h}}=\frac{H^2}{2k^3}|_{a_e},~\mathrm{where}~[\tilde{h},\pi_{\tilde{h}}]=i\hbar
\label{tensorSpectra}
\eeq
Obviously, this will lead to a exactly scale free power spectrum $k^3P_h$ for tensor perturbations. Scale dependence would occur in non-exponential inflations. In the delayed model, this could be implemented by choosing a slightly different form of pre-inflation scale factors, for example $a(0<t<\tau)\propto t^\alpha$, $\alpha\neq0$. However, power spectrums in such models could only calculated numerically. We will provide our results in the next section.

One may worry that our derivations \eqref{etaDefinition}-\eqref{hoSpectra} may be invalid in the delayed inflation scenario, because the era of inflation in this case is too short to allow the quantum fluctuation of space-time to evolve and grow out of the horizon. However, we note that even in the standard inflation scenario, fluctuations are passively stretched  instead of actively growing outside the horizon. So the question of some fluctuation mode could or not evolve out of the horizon is not determined by the absolute time inflation lasts, but  by the number of e-foldings inflation accomplishes. As long as enough number of e-foldings could be achieved, the mode's being stretched out of horizon is a indispensable phenomena.  

For scalar perturbations, paralleling with equation \eqref{hhTilde}, we have to define and quantize the auxiliary variable $\tilde{\Phi}$
\beq{}
\tilde\Phi\equiv a^{\frac{3}{2}(1+c_s^2)}\Phi=\hat{b}_{\vec{k}}u(\vec{k},\eta)+\hat{b}^\dag_{\vec{k}}u(\vec{k},\eta)
\label{varphiDefinition}
\eeq
\beq{}
u''+[c_s^2k^2+\frac{1-3c_s^2}{2}\frac{a''}{a}-\frac{9c_s^4+7}{4}\mathcal{H}^2]u=0
\label{scalarpertub}.
\eeq
In exponential inflations following from pre-inflation form of scale factor $a\propto t^0$, this equation for $\tilde{\Phi}$ has almost the same form as that of $\tilde{h}$, thus similar solution and variances
\bea{}
&&\hspace{-5mm}
P_{\tilde{\Phi}}\equiv|u(\vec{k},\eta)|^2
\xrightarrow{-k\eta\ll1}-\frac{\eta}{4\pi}\big[\Gamma(\nu)\big(-\frac{c_sk\eta}{2}\big)^{-\nu}\big]^{2}
\\
&&\hspace{-5mm}{\nu^2} \equiv \frac{9}{4}{\big(c_s^2+\frac{2}{3}\big)^2}
\eea
Using this result and equation \eqref{varphiDefinition}, we can write down the power spectrum of scalar perturbation $\Phi$ 
\beq{}
{P_\Phi }(k)|_{a_e}\equiv\frac{1}{a_e^{3(1+c_s^2)}}\frac{\Gamma^2(\nu)(-\eta)^{1-2\nu}}{4\pi(c_sk/2)^{2\nu}}|_{a_e}
\xrightarrow{c_s^2=1/3}\frac{1}{a_e^2}\frac{H^2}{2(c_sk)^{3}}|_{a_e}
\label{PPhi}
\eeq
Similar to tensor perturbations, this is again a scale free as long as $c_s^2=1/3$. This implies us that, in the delayed inflation, a radiation dominated beginning stage is completely enough to produce an exact scale free power spectrum. While a non totally radiation dominated beginning or a non-exponential inflation could both introduce the scale-dependence of the power spectrum. We will use numeric method to study such cases.

From equations \eqref{tensorSpectra} and \eqref{PPhi}, we can easily calculate the tensor-to-scalar ratio of the power spectrum,
\beq{}
r=\frac{P_h}{P_\Phi}(\alpha,w,H_i\tau,k/H_i)\stackrel{\alpha=0,c_s^2=1/3}{=\!=}3^{-\frac{3}{2}}a_e^2=e^{2H_i\tau}
\eeq
Obviously, for non-exponential type inflation, this ratio is very difficult to calculate analytically. However, for exponential type inflations, this ratio just equal to the scale factor of the universe as inflation ends. This may look strange at first glance. However, this is similar to the fact that $r$ is function of slow roll parameters, for instance $r=12.4\epsilon$ of \cite{LLReview}. The difference is that, in such scalar field driving models, $\epsilon$'s dependence on $a_e$ is hidden. In the delayed model, $e^{H_i\tau}$ plays just the same role as $\epsilon$ does in scalar field driving models.

From analysis in this section, we can see that in the delayed inflation model, the form of pre-inflation scale factor function plays the same role as the form of potentials controlling the scalar field evolution. However, for general form for pre-inflation scale factors $a(0<t<\tau)$, we have to use numerics to calculate function form of both the background scale factors $a(\tau<t)$ and perturbation quantities $\Phi$ and $h$. 

\subsection{Non-exponential inflation, numeric results}
\begin{figure*}[t]
\setcaptionwidth{0.85\textwidth}
\begin{center}
\includegraphics[scale=0.5]{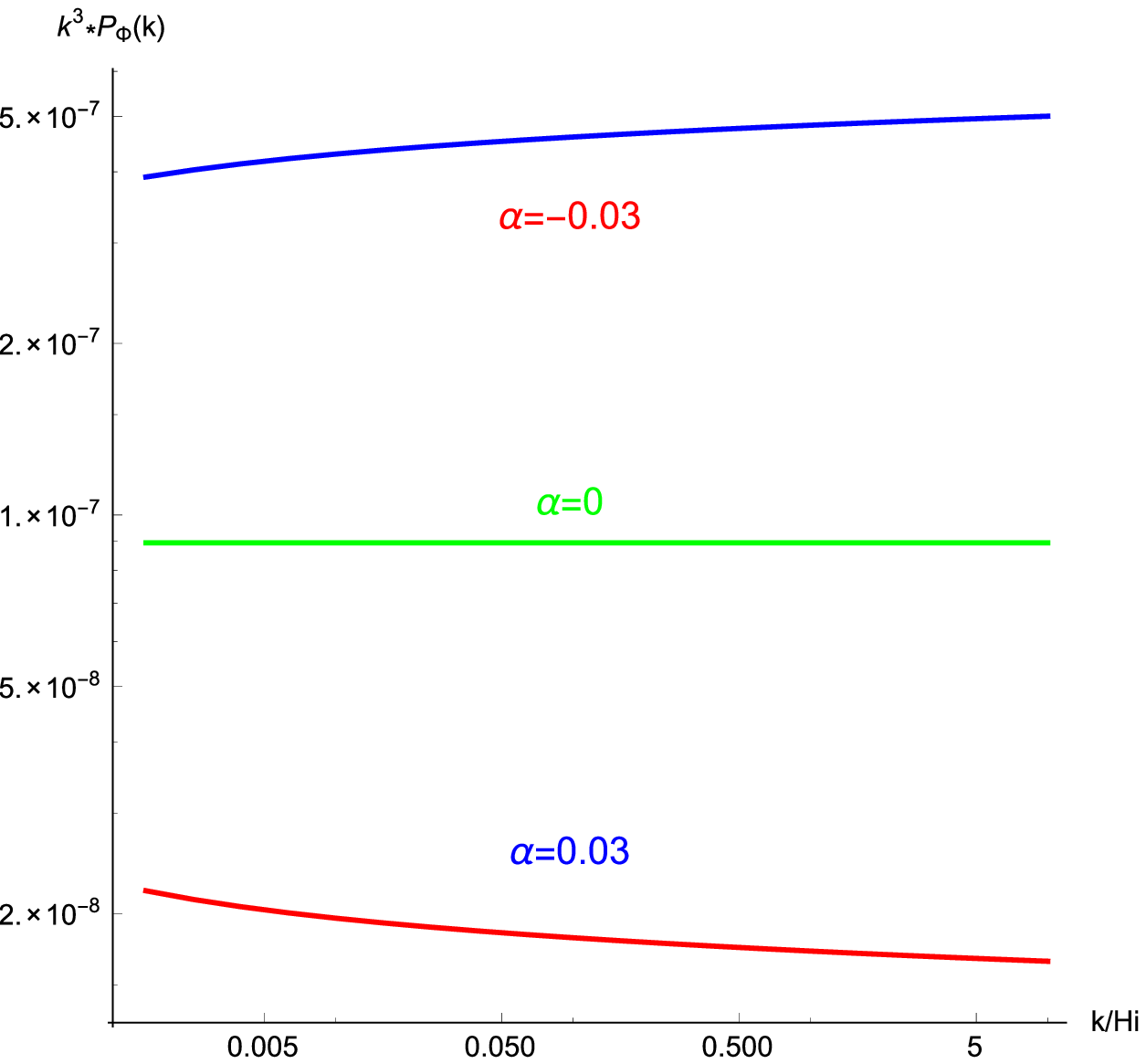}
\includegraphics[scale=0.5]{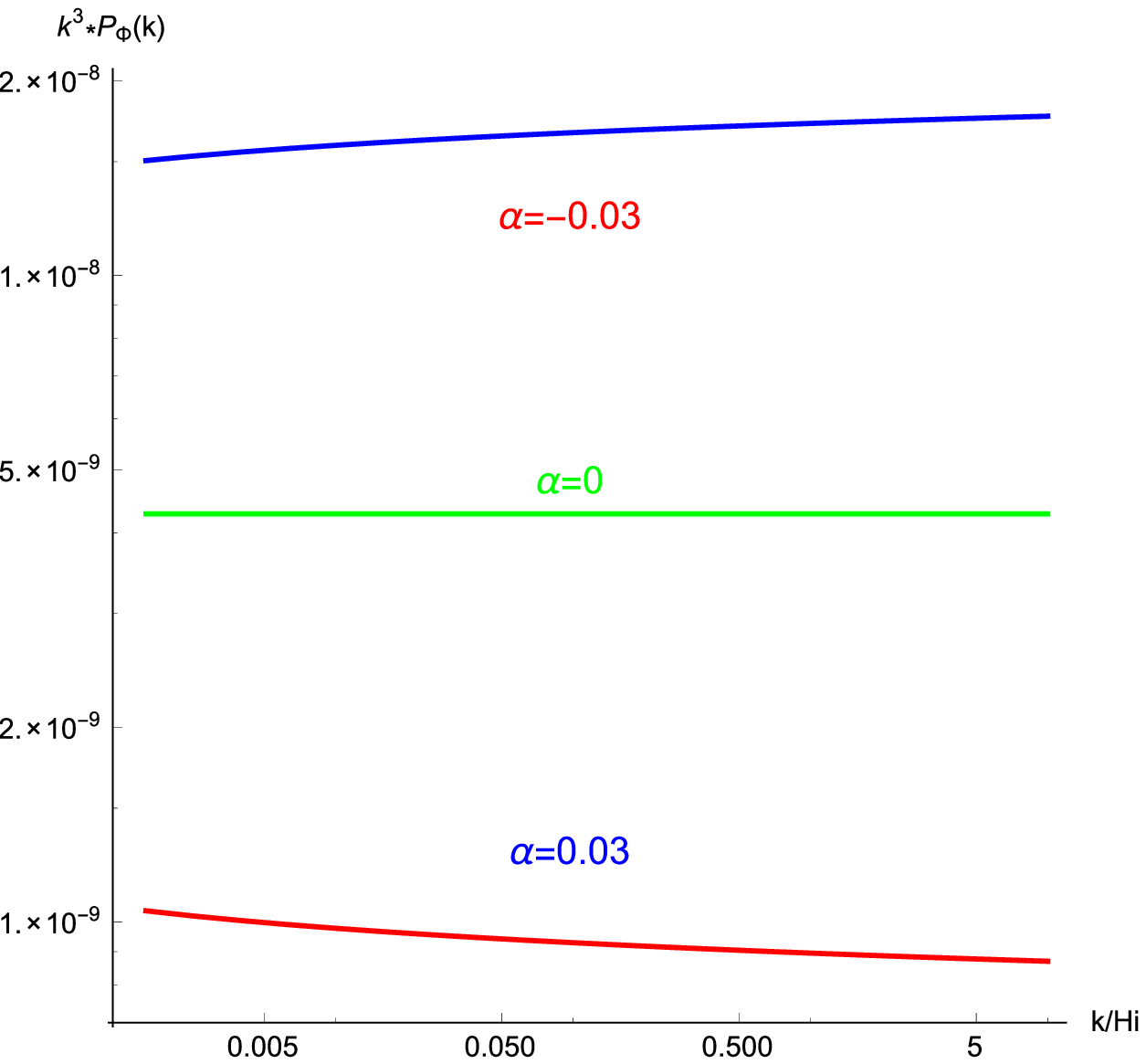}
\caption{The power spectrum of scalar (left) and tensor (right) perturbations in the delayed inflation with pre-inflation scale factor $a(0<t\leqslant\tau)\sim t^\alpha$, $\alpha=0$(green),$\pm0.03$(red/blue). }
\label{delayedSspectrum}
\end{center}
\end{figure*}

For the special form of pre-inflation scale factor $a(0<t<\tau)\propto t^\alpha$, the succeeding evolution of $a(\tau<t<2\tau)$ is analytically known, see equation \eqref{aoftDelayed}.  For simplicity we will take this form as a basic input and pay our attention in this section on the calculation of power spectrums on this background. We use numeric method solving equations \eqref{eqTildeh} and \eqref{varphiDefinition} with boundary conditions
\beq{}
v(k,\eta)_{-k\eta\gg1}=\frac{1}{\sqrt{2k}}e^{-ik\eta},~u(k,\eta)_{-k\eta\gg1}=\frac{1}{\sqrt{2c_sk}}e^{-ic_sk\eta}
\eeq
and use definitions \eqref{tensorSpectra} and \eqref{PPhi} to calculate the corresponding power spectrums. Techniquely, we use two strategies to implement high precisions. The first is, we factorize the variables $v$ and $u$ into the product of two parts, one of which is the exact solution to the corresponding differential equation when $\alpha=0$. In more details, we let 
\beq{}
v(k\eta)_{\alpha\neq0}=\frac{1}{\sqrt{2k}}e^{-ik\eta}(1-\frac{i}{k\eta})C(k\eta)
,~u(c_sk\eta)_{\alpha\neq0}=H^{(1)}_{\nu}(c_sk\eta)D(c_sk\eta)
\eeq
and pay the main attentions on the functions $C(k\eta)$ and $D(c_sk\eta)$. Since the analytically part carries out the early($-k\eta\gg1$)-vibrational and late($-k\eta\ll1$)-exponentially-growing part, the numerical solving of $C$ and $D$ becomes more viable for high precision implementation. The second is, we take the cosmic time $t$ instead of the conformal time $\eta$ as the independent variables. Because during the inflationary era, $\tau\approx-\exp(t)$. $\tau$ varies in a exponentially large range while $t$ varies only in a linearly increasing range. Obviously, for non-vibrational and non-exponentially-growing functions such as $C$ and $D$, following their evolution on the $t$ axis is more viable than on the $\tau$ axis.

Our results are displayed in Figure \ref{delayedSspectrum}. From the figure, we easily see that, the parameter $\alpha$ in $a(0<t<\tau)\propto t^\alpha$ is directly related with the scale dependence of power spectrum of perturbations. When $\alpha>0$, the power spectrum is slightly red-tilt. While as $\alpha<$ the spectrum is purely blue-tilt.  Just as we mentioned above, the role of $a(t)_{0<t<\tau}$'s form in the delayed model is totally equivalent with that of potentials in the scalar field driving inflation.

\begin{figure*}[hb]
\setcaptionwidth{0.85\textwidth}
\begin{center}
\includegraphics[scale=0.9]{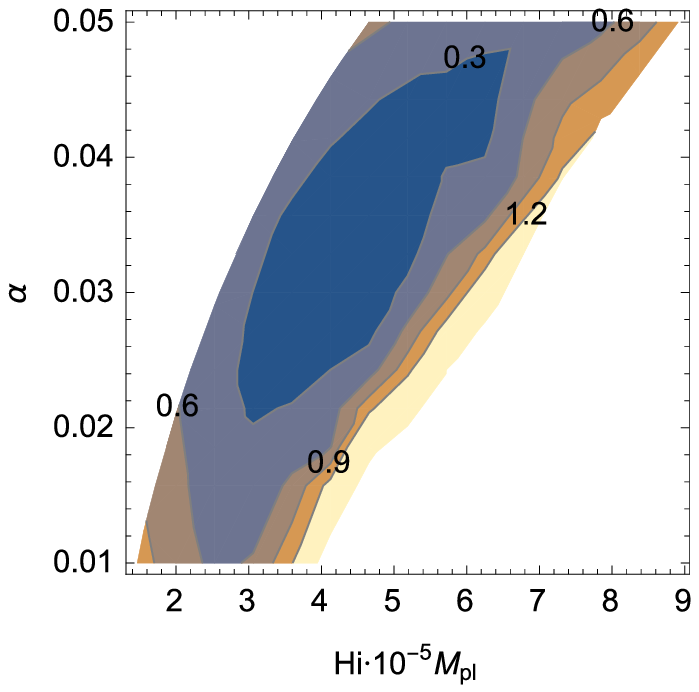}
\includegraphics[scale=0.9]{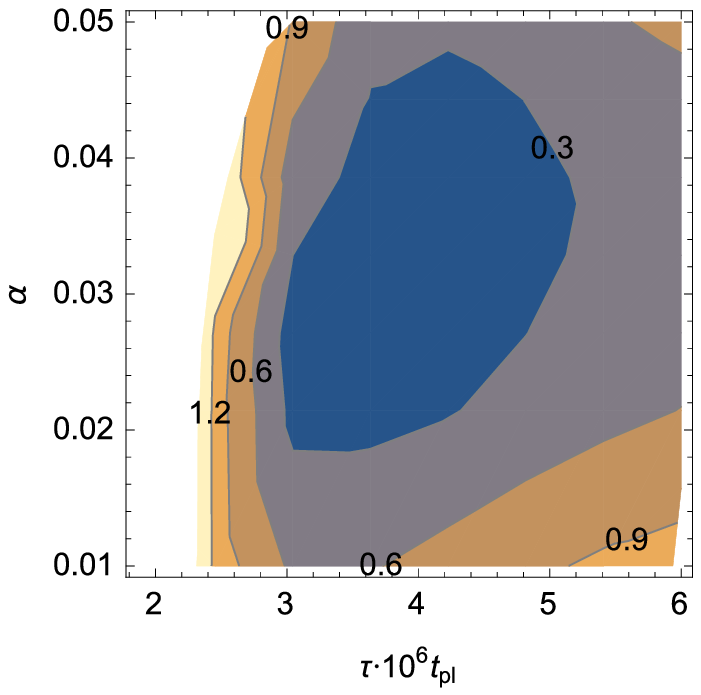}
\caption{The distribution of error function \eqref{errFunc} on the $H_i$-$\alpha$ and $\tau$-$\alpha$ plane.}
\label{figFITparameter}
\end{center}
\end{figure*}
As long as the relevant parameter are concerned in figure \ref{delayedSspectrum}, we first recall that in Ref. \cite{delayedCosmology}, Choudhury et al discussed the constraints on $\alpha$ and $\tau$ imposed by the number of e-foldings $65\leqslant N_e$. Their conclusion is, $\tau$ is of order $10^2\sim10^3t_{pl}$, $\alpha$ could be negative as well as positive in a rather broad range, see the Figure 2 of their paper. However, this conclusion is based on the condition that $H_i=1$ (in their original notations, this is $\rho_0^{1/4}$), whose default unit is $M_{pl}$. In our paper we will relax this pre-assumption and take $H_i$ as free parameter to determine. We will use the following conditions
\begin{itemize}
\item number of e-foldings $\frac{a_e}{a_\mathrm{begin}}=\exp[H_i\frac{\tau^{1-\frac{3}{2}(1+w)\alpha}}{1-\frac{3}{2}(1+w)\alpha}]=\exp(65)$ \cite{LLReview}
\item COBE normalization $k^3P_\Phi=\frac{50\pi^2}{9}\cdot(1.9\times10^{-5})^2=1.979\times10^{-8}$ \cite{WMAP}
\item power spectrum index $n_s=0.9677-1$ of scalar perturbations \cite{PLANCK2015I}
\item Planck+BICEP2 upper bound on tensor to scalar ratios $\frac{P_h}{P_\Phi}\leq0.12$ \cite{BICEP2_PLANCK}, we will take $\frac{P_h}{P_\Phi}=0.01$ as examples
\end{itemize}
to define error function
\beq{}
errFunc=\big(\frac{k^3P_\Phi|_{k_{COBE}}}{1.979\times10^{-8}}-1\big)^2+\big(\frac{n_s}{0.0323}-1\big)^2+\big(\frac{P_h/P_\Phi}{0.01}-1\big)^2
\label{errFunc}
\eeq
Under the extra assumption that $c_s^2=\frac{\partial p}{\partial\rho}|_{adiabatic}=\frac{1}{3}$, we find that the following suit of parameters
\beq{}
H_i=4.62\times10^{-5}M_\mathrm{pl},~\tau=3.59\times10^6 t_\mathrm{pl},~\alpha=0.0334,~a_e=0.0437
\eeq
minimises the above error function. Centering on this point, the distribution of the error functions on the $H_i-\alpha$ and $\tau-\alpha$ plane is displayed  in figure \ref{figFITparameter}. From this distribution, we easily see that in the delayed model,  inflation occurs at energy scale of $10^{-5}M_{pl}$ while the time delaying is of order $10^6t_{pl}$.

\section{Late time evolutions}
\label{sectionLateTimeEvolution}
\begin{figure*}[b]
\setcaptionwidth{0.85\textwidth}
\begin{center}
\includegraphics[scale=0.8]{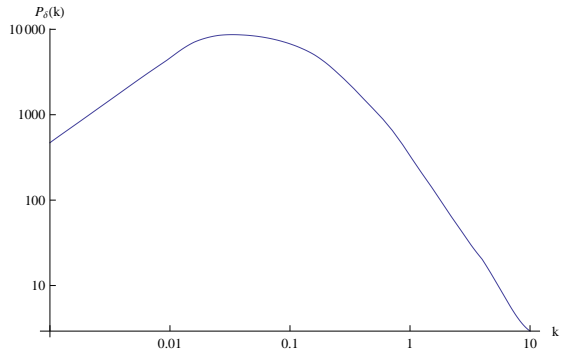}
\caption{Power spectrum of matter distributions predicted by the delayed inflation and delayed $\Lambda$CDM cosmology with delaying $10^6t_\mathrm{pl}$. The magnitude and shape of this late time power spectrum is very close to observations provided in \cite{2dF, SDSS1, SDSS2}.}
\label{todayPowerMatter}
\end{center}
\end{figure*}
Some people may worry that delays as large as $\tau\sim10^6t_{pl}$ may cause noticeable effects on the late time evolution of the universe, for instance, the structure formation processes. However, from the viewpoint of post-inflationary cosmologies, this is in fact an almost negligible period of time. To show that this is the case indeed, let us consider the evolution of matter/energy fluctuations in the late time of $\Lambda$- Cold Dark Matter model in this section. The basic idea of this section is just an exercise from \cite{ModernCosmology} and is essentially using numerics to search solutions of the following equation array
\begin {equation}
\Theta_0 '+ \frac{k}{3}{\Theta _1} =  -\Phi',
\label{0moment}
\end{equation}
\begin {equation}
\Theta_1 '+ \frac{k}{3}{\Theta _0} =  - \frac{k}{3}\Phi,
\end{equation}
\begin {equation}
\delta'  + ikv =  - 3\Phi,
\end{equation}
\begin {equation}
v' + \frac{{\dot a}}{a}v = ik\Phi,
\end{equation}
where $\Theta_0$ and $\Theta_1$ are the zeroth and first moment of cosmic background radiation, $\delta$ and $v$ are those of matters, including both dark matter and baryons, $\Phi$, $k$ and $a$ are respectively scalar-perturbation, wave number and scale factors of the late time universe. $\Phi$ and $a$ satisfy equations
\beq{}
{k^2}\Phi  + 3\mathcal{H}\left( {\Phi'  + \mathcal{H}\Phi } \right) = \frac{1}{2}{a^2}\left[ {\rho \delta  + 4{\rho _r}{\Theta _0}} \right]
\label{00scalarPertub} 
\eeq
\beq{}
\big[\frac{\dot a(t)}{a(t)}\big]^2=\frac{H_0^2}{3}\big[\frac{\rho_m(t-\tau)}{\rho_\mathrm{tot}}+\frac{\rho_r(t-\tau)}{\rho_\mathrm{tot}}+\frac{\rho_\Lambda}{\rho_\mathrm{tot}}\big]
\label{lateFriedman}
\eeq
Note again that in these equations, overdot represent derivatives respective to the physical time $\frac{d}{dt}\equiv\frac{d}{ad\eta}$, while $^\prime$, to the conformal time $\frac{d}{d\eta}$.

With the early time approximation and the primordial power spectrum of figure \eqref{delayedSspectrum} as input
\beq{}
\Phi(\eta_i)=2\Theta_0=\sqrt{P_\Phi}
\label{thetaini}
\eeq
\beq{}
\delta(\eta_i)=3\Theta_0
\eeq
\beq{}
\Theta_1(\eta_i)=0
\eeq
\beq{}
v(\eta_i)=0
\label{vini}
\eeq
we get from Eqs \eqref{0moment}-\eqref{lateFriedman} the power spectrum of matters in the universe today, $P_{\delta}\equiv|\delta_k|^2_\mathrm{today}$. The result is illustrated in Figure \ref{todayPowerMatter}. From the figure we easily see that, delays as long as $10^6t_\mathrm{pl}$, introduces almost no observable effects on structure formation and evolutions in the late universe

\section{Conclusion}
\label{sectionConclusion}
This paper calculated the power spectrum of primordial perturbations, including both scalar and tensor types, produced during the inflations driven by delays. The result indicate that, in the delayed inflation without inflaton field, a radiation dominated early universe are completely enough to provide near scale-free power spectrums for both tensor and scalar perturbations. The form of pre-inflation scale factors $a(0<t<\tau)$ plays the same role as the potential of scalar fields in conventional inflations. And deviations from a pure radiation dominating inflationary universe, or the adjustment of model parameter $\alpha$ may both introduce the dependence on scales for the power spectrum. Using observations of Planck and BICEP2, we estimated the key parameters of the model. $\tau\approx10^{6}t_{pl}$, $H_i=H_e\approx10^{-5}M_{pl}$. In the last section, we add delays to the $\Lambda$CDM and numerically evolve the perturbations in the late time universe and get the power spectrum of matters in the current universe. The result is also consistent with the standard cosmological model and observations.

As discussions, we comment here that, i) if inflations are really caused by delay, then we need not any inflaton fields or exotic dark energies to dominate the early universe. The big bang just begins from the radiation time. As a result, we also need no mechanism of reheating. This is indeed a simplification of the early universe. ii) although the delay mechanism could drive the early time inflation, it could not alleviate our thirsty of dark energies in the late time universe.
iii) in the delayed inflation, the scale factor $a$ is continuous but not smooth at the point $t=\tau$. Revealing effects of this feature on observations is a very valuable work for futures

\section*{Acknowledgements}
This work is supported by Beijing Municipal Natural Science Foundation, Grant. No. Z2006015201001.

\end{CJK}

\begin{thebibliography}{99}

 \bibitem{Guth}
A. Guth,
"The Inflationary Universe: A Possible Solution to the Horizon and Flatness Problems", 
 {\em Phys. Lett. B} {\bf 108} (1982) 389.
 
 \bibitem{Linde}
A. D. Linde,
“A New Inflationary Universe Scenario: A Possible Solution of the Horizon, Flatness, Homogeneity, Isotropy and Primordial Monopolo Problems",
 {\em Phys. Lett. B} {\bf 108} (1982) 389.
 
 \bibitem{LLReview}
A. Liddle and D. Lyth,
“Cosmological inflation and large-scale structure",
 \href{http://inspirehep.net/record/533080/?ln=zh_CN}{{\tt INSPIRE:533080}}.
 
 \bibitem{BaumannReview}
Daniel Baumann
 ''TASI Lectures on Inflation'',
 \href{http://arxiv.org/abs/0907.5424}{{\tt arXiv:0907.5424}}.\\
 Daniel Baumann, Liam McAllister
 ''Inflation and String Theory'',
  \href{http://arxiv.org/abs/1404.2601}{{\tt arXiv:1404.2601}}.
 
  \bibitem{WMAP}
G.Hinshaw et al,
“Nine-Year Wilkinson Microwave Anisotropy Probe (WMAP) Observations: Cosmological Parameter Results", 2012
 \href{http://arxiv.org/abs/1212.5226}{{\tt arXiv:1212.5226}}.
 
 \bibitem{BICEP2}
 BICEP2 Collaboration (Ade, P.A.R. et al.),
 ''Detection of B-Mode Polarization at Degree Angular Scales by BICEP2'',
 {\em Phys.Rev.Lett. }{\bf 112}  (2014) 241101,
 \href{http://arxiv.org/abs/arXiv:1403.3985}{{\tt arXiv:1403.3985}}.
 
  \bibitem{BICEP2_PLANCK}
 BICEP2/Keck, Planck Collaborations
 ''A Joint Analysis of BICEP2/Keck Array and Planck Data'',
 {\em Phys.Rev.Lett. }
 \href{http://arxiv.org/abs/1502.00612}{{\tt arXiv:1502.00612}}.

 
 \bibitem{delayedCosmology}
 D. Choudhury, D. Ghoshal, A. A. Sen,
 ''Standard Cosmology Delayed'',
 {\em JCAP} {\bf 02} (2012) 046,
 \href{http://arxiv.org/abs/1106.6231}{{\tt arXiv:1106.6231}}.
 
  \bibitem{CPT}
P. Peter,
 ''Cosmological Perturbation Theory'',
Lecture notes from the Mangaratiba cosmology school, August 2012,
 \href{http://arxiv.org/abs/1303.2509v2}{{\tt arXiv:1303.2509}}.


 \bibitem{CIP}
W. H. Kinney
 "Cosmology, inflation, and the physics of nothing",
 {\em NATO Sci. Ser. II}  {\bf123}(2003) 189,
 \href{http://arxiv.org/abs/astro-ph/0301448v2}{{\tt arXiv:0301448}}.
 
 \bibitem{ModernCosmology}
 S. Dodelson,
 ''Modern 	Cosmology'',
 Elsevier Pte Ltd., 2008 version

\bibitem{spectrum}
E.D.Stewart, D.H.Lyth,
 ''A more accurate analytic calculation of the spectrum of cosmological perturbations produced during inflation'',
 {\em Phys.Lett.B, }{\bf302} (1993) 171-175,
 \href{http://arxiv.org/abs/gr-qc/9302019v1}{{\tt arXiv:9302019v1}}.
    
 \bibitem{Planck}
 Planck Collaboration  (P. A. R. Ade et al.)
  "Planck 2013 results. XXII. Constraints on inflation",
   {\em Astron. Astrophys. }{\bf 571}  (2014) A22,
 \href{http://arxiv.org/abs/1303.5082}{{\tt arXiv:1303.5082}}.

 \bibitem{COBE}
 Bennett et al.
  "4-Year COBE DMR Cosmic Microwave Background Observations: Maps and Basic Results",
   {\em Astrophys.J }{\bf 464:}  (1996) L1-L4,
 \href{http://arxiv.org/abs/1303.5082}{{\tt arXiv:1303.5082}}.
 
  \bibitem{2dF}
Will J. Percival et al.
 ''The 2dF Galaxy Redshift Survey: The power spectrum and the matter content of the universe'',
{\em Mon.Not.Roy.Astron.Soc}(2001) 327:1297,
 \href{http://arxiv.org/abs/astro-ph/0105252}{{\tt arXiv:0105252}}.
 
   \bibitem{SDSS1}
M Tegmark et al.
 ''The 3D power spectrum of galaxies from the SDSS'',
{\em Astrophys.J}{\bf606}(2004) 702-740,
 \href{http://arxiv.org/abs/astro-ph/0310725v2}{{\tt arXiv:0310725}}.
 
    \bibitem{SDSS2}
D. J. Eisenstein et al.
 'Detection of the Baryon Acoustic Peak in the Large-Scale Correlation Function of SDSS Luminous Red Galaxies'',
{\em Astrophys.J}{\bf633}(2005) 560-574,
 \href{http://arxiv.org/abs/astro-ph/0501171v1}{{\tt arXiv:0501171}}.
 
 \bibitem{PLANCK2015I}
 Planck Collaboration,
 ''Planck 2015 results. I. Overview of products and scientific results'',
 {\em Astro. \& Astrophys. }
 \href{http://arxiv.org/abs/1502.01582}{{\tt arXiv:1502.01582}}.
 
\end{thebibliography}
\end{document}